\documentclass[aps,prb,twocolumn,groupedaddress]{revtex4}

\usepackage{epsfig}
\usepackage{graphics}

\begin{document}
% \draft command makes pacs numbers print
\draft

\title{Ultrafast carrier relaxation in GaN, In$_{0.05}$Ga$_{0.95}$N and an
In$_{0.05}$Ga$_{0.95}$N/In$_{0.15}$Ga$_{0.85}$N Multiple Quantum Well}

\author{\"{U}mit \surname{\"{O}zg\"{u}r}, and Henry O. \surname{Everitt}}
\email[Email address]{: everitt@aro.arl.army.mil}

\affiliation{Department of Physics,Duke University, Durham, NC 27708}

\date{\today}

\begin{abstract}

Room temperature, wavelength non-degenerate ultrafast pump/probe
measurements were performed on GaN and InGaN epilayers and an InGaN
multiple quantum well structure. Carrier relaxation dynamics were
investigated as a function of excitation wavelength and intensity.
Spectrally-resolved sub-picosecond relaxation due to carrier
redistribution and QW capture was found to depend sensitively on the
wavelength of pump excitation.  Moreover, for pump intensities above a
threshold of 100 $\mu$J/cm$^2$, all samples demonstrated an additional
emission feature arising from stimulated emission (SE).  SE is evidenced
as accelerated relaxation ($<$ 10 ps) in the pump-probe data,
fundamentally altering the re-distribution of carriers.  Once SE and
carrier redistribution is completed, a slower relaxation of up to 1 ns for
GaN and InGaN epilayers, and 660 ps for the MQW sample, indicates carrier
recombination through spontaneous emission.

\end{abstract}


\maketitle

\section{Introduction}

Technological advances in group-III nitride-based optoelectronics have
been possible with extensive materials research, resulting in the
commercialization of short wavelength emitters and
detectors.\cite{NakamuraBLD1997,NakamuraJJAP95,NakamuraJJAP1997-2,CarranoJAP1998}
The active layers in high efficiency emitters, such as blue/green light
emitting diodes and blue/purple laser diodes, contain InGaN alloys. Time
scales for carrier recombination, transport, and quantum well capture in
the ultrafast regime determine the efficiency of optoelectronic devices.
Therefore it is important to understand the carrier relaxation and
recombination mechanisms in InGaN structures. In addition to many studies
on recombination
times,\cite{ChichibuJVST1998,ChichibuMSEB1999,BerkowiczPRB2000} there have
been a limited number of reports on ultrafast carrier dynamics in InGaN
heterostructures\cite{ChoiPRB2001-2,OmaePSSA2002} and multiple quantum
wells (MQW).\cite{KawakamiAPL2000,SatakePRB1999} Measurements on
heterostructures, single QWs, and MQWs have emphasized different aspects
of carrier relaxation in nitrides. In this paper, we report comprehensive
room temperature ultrafast measurements on a bulk GaN, an InGaN epilayer,
and an InGaN MQW sample having barriers with the same In composition as
the epilayer sample.

In achieving the current state of nitride device development, overcoming
material growth difficulties has been the main focus. The efficiencies of
InGaN-based emitters are strongly affected by material inhomogeneities
such as compositional fluctuations and indium-phase separation. However,
inhomogeneities in the form of quantum dot-sized In-rich regions are
observed to increase lateral confinement, thereby increasing the emission
efficiencies of InGaN devices. The effect of these In-rich regions on
electron-hole recombination through spontaneous emission (SPE) has been
studied extensively.\cite{ChichibuJVST1998,MartinAPL1999,RibletAPL1999} In
spite of imperfect material properties, high emission efficiencies are
observed not only through enhanced SPE but also through stimulated
emission (SE) generated at moderate pump powers. There have been many
reports on SE in InGaN epilayers\cite{SatakePRB1998,ShmaginJAP1997}and MQW
structures.\cite{SchmidtAPL1998,SchmidtAPL1998-2,BidnykAPL1998} However SE
and its effects on carrier relaxation is poorly understood. In this study,
ultrafast dynamics in the presence and absence of SE are investigated.

Three samples, a GaN epilayer, an InGaN epilayer, and an InGaN MQW
structure, were grown on c-plane double polished sapphire by metalorganic
chemical vapor deposition at the University of California, Santa
Barbara.\cite{KellerJCG1998} The GaN sample is $\sim$ 3 $\mu$m thick, and
the buffer layers in the other two samples are $\sim$ 2 $\mu$m thick
GaN:Si. InGaN epilayer sample consists of a 60 nm thick
In$_{0.05}$Ga$_{0.95}$N:Si layer, capped with another 15 nm thick GaN
layer. The InGaN MQW is a typical laser active layer structure with 10
periods of 8.5 nm In$_{0.05}$Ga$_{0.95}$N:Si barriers and 3.5 nm
In$_{0.15}$Ga$_{0.85}$N quantum wells. There is a 100 nm GaN cap layer on
top of the MQW. The doping in the barriers and in the InGaN epilayer is
$\sim$ 10$^{18}$ cm$^{-3}$.

In section II, techniques used in the experiment are introduced. Then the
results for each sample are respectively considered as subsections of
section III. In each subsection, first continuous-wave (cw) and time
integrated measurements are presented to identify the band structures. The
investigation of carrier relaxation dynamics follows, using non-degenerate
time-resolved differential transmission (TRDT) spectroscopy. To understand
the effects of the underlying buffer and cap GaN layers, and to understand
the effects of the 3D InGaN barriers in the MQW sample, measurements on
the bulk GaN sample and on the InGaN epilayer sample are discussed,
respectively.  Above- and below-bandgap excitation data are presented to
explore the excitation wavelength dependence of the GaN and InGaN
relaxation phenomena.

Similarly, measurements on the MQW sample are reported for above-, at-,
and below-barrier energy excitations. Time-resolved data for the MQW
sample are compared with results from a previous study of carrier capture
times using degenerate TRDT.\cite{OzgurAPL2000}

\section{Experimental Techniques}

\subsection{Continuous wave and time-integrated characterization}

Continuous-wave (cw) photoluminescence (PL), PL excitation (PLE), and
absorption measurements were performed at room temperature. Cw-PL was
measured at excitation power densities of 100 W/cm$^2$ using a 25 mW HeCd
laser operating at 325 nm (3.82 eV). A 300 W Xe Lamp was used for PLE and
cw-absorption measurements. Xe Lamp was dispersed by a 30 cm double
grating spectrometer for the PLE, and both cw-PL and PLE were detected by
a photomultiplier tube attached to a 75 cm grating spectrometer. Another
30 cm grating spectrometer with a charge coupled device (CCD) was used for
the absorption measurements.

Pulsed excitation, time-integrated PL (TI-PL) is performed on the samples
at room temperature using 1 kHz, $\sim$10 $\mu$J pulses from an optical
parametric amplifier, at excitation power densities varied between 20
$\mu$J/cm$^{2}$ and 2 mJ/cm$^{2}$, either by changing the focus or using
neutral density filters. For high enough excitation densities, SE features
are observed for all the samples. To obtain the SE threshold densities, PL
from the samples was detected both normal to the surface from the front
and parallel to the surface from the edges. Edge detection produced
stronger SE signals, making it the preferred detection scheme. The samples
were excited normal to the surface, and the PL was collected from the edge
using a 600 $\mu$m diameter UV-VIS fiber. A CCD attached to the 30 cm
spectrometer was used for detection. Both the spectrally-integrated PL
intensity and the emission linewidth were plotted as a function of pump
power density and found to give identical SE thresholds.

\subsection{Time-resolved characterization}

Previously, standard wavelength-degenerate, time-resolved differential
transmission (TRDT) measurements were performed on the MQW sample, using a
frequency-doubled mode-locked Ti:Sapphire laser. \cite{OzgurAPL2000} In
this study, non-degenerate TRDT spectroscopy was applied at room
temperature. A Ti:Sapphire laser-seeded, 1 kHz Quantronix Titan
regenerative amplifier (RGA) with 1.8 mJ , 100 fs pulses at 800 nm was
used. Half of the RGA output power is used to pump a Quantronix TOPAS
optical parametric amplifier (OPA). The signal output from the tunable OPA
is frequency quadrupled and used as the pump in the DT experiment. The
other half of the RGA output is frequency doubled in a BBO crystal and
focused on a quartz cell filled with D$_2$O to generate a broadband
continuum probe centered near 400 nm (3.11 eV). The continuum probe, which
has relatively minor amplitude fluctuations over its $>$ 100 nm bandwidth,
is highly attenuated by spatial filtering, then collected and focused on
the sample using a spherical mirror. The pump beam is delayed with respect
to the probe beam using a retroreflector mounted on a 1 $\mu$m-resolution
translational stage. The probe, transmitted through the sample, is then
collected by a UV liquid light guide, sent to the 30 cm grating
spectrometer, and detected by a CCD camera attached to the output port. It
is important to note that the excitation regime in this study is much
higher ($>$100 $\mu$J/cm$^2$) than the previous study of degenerate TRDT
($\sim$1 $\mu$J/cm$^2$).\cite{OzgurAPL2000} Using a pump
intensity-independent absorption constant of 10$^5$ cm$^{-1}$,
\cite{MuthAPL1997} the number of photo-injected carriers in the MQW sample
is estimated to be 10$^{17}$ cm$^{-3}$ for the previous study, and
$>$10$^{19}$ cm$^{-3}$ for this work. These carrier densities are larger
for the GaN sample due to its larger thickness.

First, the continuum probe spectrum is recorded.  Next, the pulsed
absorption spectrum of the sample is obtained by comparing the continuum
probe spectrum and the spectrum of the probe transmitted through the
unpumped sample. Some absorption features that were not clear in the
cw-absorption data were easily observed in the pulsed absorption spectra
due to larger excitation density. Then, pump-probe measurements are made
by comparing the transmission of the probe through the sample with and
without the pump beam for various delays up to 0.4 ns. The absolute DT
signal at energy h$\nu$ is
\begin{equation}
DT(h\nu)=\frac{T_{PumpON}(h\nu)-T_{PumpOFF}(h\nu)}{T_{PumpOFF}(h\nu)}=
-\triangle\alpha(h\nu) d,
\end{equation}
where $T_{PumpON}$ and $T_{PumpOFF}$ are the probe transmission signal
magnitudes with the pump beam on and off, respectively. $\triangle \alpha$
is the change in the absorption coefficient, and $d$ is the thickness of
the sample over which the change in absorption is induced. Pump and probe
spot diameters on the samples were $\sim$1 mm and $\sim$0.2 mm,
respectively. The absorption from the bare probe beam was observed to be
much smaller than the absorption due to the pump beam, suggesting that the
modulation in the DT signal is purely from the pump.

TRDT data is presented in two different ways, spectrally-integrated and
spectrally-resolved, in order to elucidate different aspects of the
relaxation processes.  Spectrally-integrated DT gives an indication of the
total population of carriers (i.e. density of states) and their aggregate
decay, while spectrally-resolved DT describes the distribution of carriers
and indicates their energy relaxation pathways.

Due to the thick GaN layers, strong absorption at the GaN band edge allows
only weak transmission of the probe beam for all the samples. Therefore, a
small pump-induced change in the absorption at the GaN energy will yield a
large TRDT signal. Spectrally-resolved TRDT data for all the samples show
oscillations that arise from the interference of the multiple reflections
in the sample.

Similar non-degenerate TRDT measurements have been performed on
GaN,\cite{OmaePRB2002,ChoiPRB2001} InGaN
epilayers\cite{ChoiPRB2001-2,OmaePSSA2002}, and InGaN MQWs
\cite{KawakamiAPL2000,SatakePRB1999,OmaePSSA2002,KawakamiJPCM2001}, by
different groups. These studies provide insights on important aspects of
carrier relaxation dynamics. In this study, a more comprehensive
investigation is made, focusing on relaxation over many time scales,
sub-picosecond to nanosecond, as a function of excitation wavelength and
density. The role of stimulated emission on carrier dynamics is also
studied in both spectrally-integrated and spectrally-resolved data.

\section{Results and Discussion}

\subsection{Differential transmission of bulk GaN}

\begin{figure}
\centerline{\resizebox{8cm}{!}{\hbox{\includegraphics {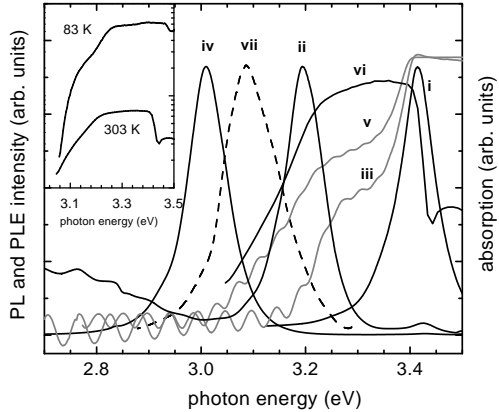}}}}
\caption{Room temperature cw-PL(i) for the GaN sample, cw-PL(ii) and
absorption(iii) for the InGaN sample, cw-PL(iv), cw absorption (v), PLE
(vi) and time integrated pulsed-PL (vii) for the MQW sample. The
excitation density for the pulsed-PL is above ($\sim$2 mJ/cm$^2$) the SE
threshold. Inset shows the log scale plot of the 83 K and 303 K PLE for
the MQW sample.} \label{MQWPLPLE}
\end{figure}

The cw-PL from the bulk GaN sample is centered at 3.41 eV
(Fig.~\ref{MQWPLPLE}) when excited above (3.82 eV) the bandgap, in
agreement with the 3.41 eV GaN band edge obtained from pulsed absorption
measurements. Pulsed-PL showed another peak at 3.31 eV due to
electron-hole plasma (EHP) induced
SE\cite{HerzogAPL2000,HolstMRSIJNSR1997,JursenasAPL2001} with a threshold
of 70 $\mu$J/cm$^2$. The SE peak is observed to redshift by 25 meV and
broaden (from 37 meV to 66 meV) with increasing pump intensity (from
40$\mu$J/cm$^2$ to 2 mJ/cm$^2$). When the sample is excited below (3.34
eV) the bandgap, no PL is observed.

\begin{figure}
\centerline{\resizebox{8cm}{!}{\hbox{\includegraphics{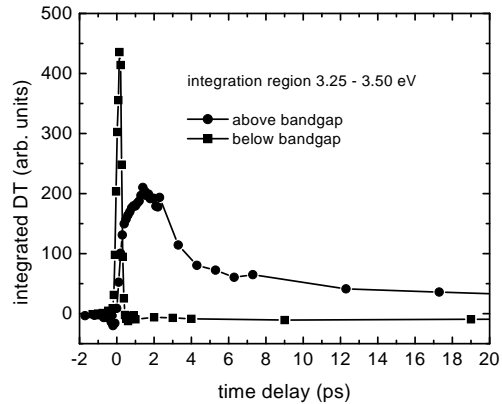}}}}
\caption{Time evolution of the spectrally-integrated TRDT for above
(circles) and below (squares) bandgap excitations of the bulk GaN sample
at an excitation power density of 300 $\mu$J/cm$^2$. Integration region
was 3.25 eV - 3.50 eV.} \label{GaNIntDT}
\end{figure}

The spectrally-integrated TRDT data for the GaN at excitation energies
above (3.82 eV) and below (3.34 eV) the bandgap are shown in
Fig.~\ref{GaNIntDT}. For below bandgap excitation, increased transmission
is observed for a very short time, less than 1 ps. Data for the above
bandgap excitation shows a similarly rapid rise ($<$ 1 ps), a $\sim$2
ps-wide peak, and a fast decay (3-6 ps), followed by a much slower
relaxation (100's of ps). For both excitations, the early, strong increase
in transmission at the GaN band edge is due to photo-excited carriers and
the dynamic (AC) Stark effect.\cite{OmaePRB2002,ChoiPRB2002} The wide peak
and fast relaxation are due to the operation and removal of carriers
through SE, which occurs when the pump density is above threshold (70
$\mu$J/cm$^2$) and ends when the number of carriers is reduced below
threshold. The time constant for SE decay at excitation density of 300
$\mu$J/cm$^2$ is measured to be $\sim$2 ps.

\begin{figure}
\centerline{\resizebox{8cm}{!}{\hbox{\includegraphics{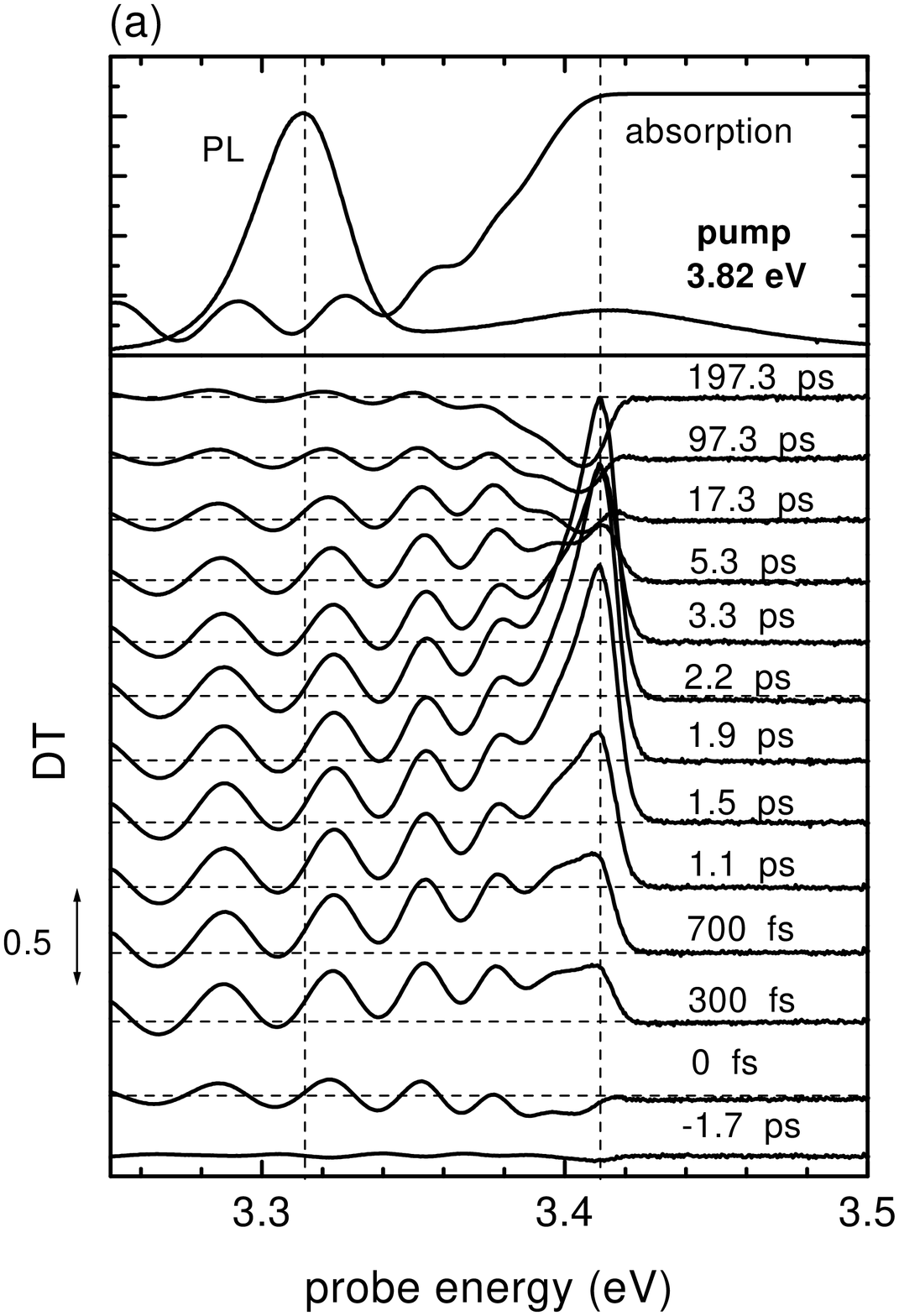}
\includegraphics{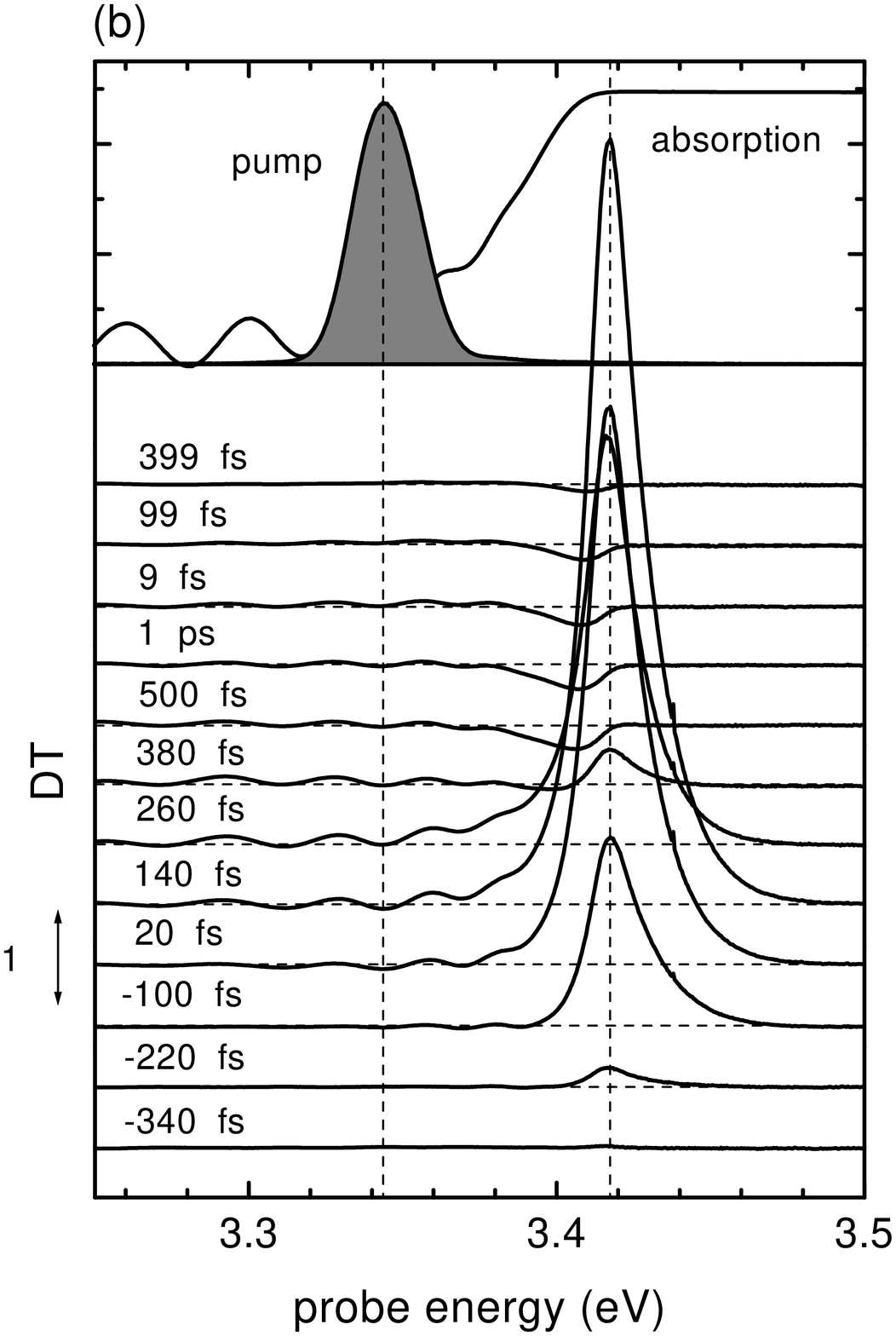}}} }
\caption{Spectrally-resolved TRDT for above (a) and below (b) bandgap
excitations of the bulk GaN sample at various delays for an excitation
density of 300 $\mu$J/cm$^2$. The top graphs show the pump spectrum, the
TI-PL, and the pulsed absorption.} \label{GaNdt}
\end{figure}

Spectrally-resolved results from the TRDT measurements on the GaN sample
(Fig.~\ref{GaNdt}) confirm these findings for excitations above and below
the GaN band edge. For above bandgap excitation (Fig.~\ref{GaNdt}a), the
carrier distribution is broad ($\sim$80 meV) and extends below the bandgap
($\sim$3.34 eV) during the first 1.2 ps.  The carriers relax to the GaN
band edge in $\sim$ 2ps through multiple LO-phonon and carrier scattering
events required to cool the carriers as much as 400
meV.\cite{GaNabsorption} The 2 ps-wide SE feature in the
spectrally-integrated DT is observed during this cooling. After 2 ps when
the carrier redistribution is finished, a clear decay due to SE becomes
visible, with a decay constant of $\sim2$ ps. At later times (17 ps), the
lower energy part (3.34 eV - 3.37 eV) of the distribution disappears,
while carriers near the GaN band edge show increasing absorption. The
decay of the low energy part might be explained by the blueshift of the
SE, and carrier energy, in the EHP after the number of carriers is
decreased through SE. This is supported by the fact that SE redshifts with
increasing carrier density due to increased coulombic
repulsion.\cite{HolstMRSIJNSR1997}

For excitation 80 meV below the bandgap (Fig.~\ref{GaNdt}b), the
generation of carriers may seem more surprising, especially because it
cannot be explained as an assist from room temperature thermal energy
k$_B$T (26 meV) or from the pulse bandwidth (25 meV). It is well known
that the excitonic resonances are visible even at room temperature for
GaN\cite{MuthAPL1997}, and AC Stark effect has been observed for GaN for
detunings as large as 159 meV below the excitonic
resonance.\cite{ChoiPRB2002} The bleaching which exists during the intense
pump pulse and lasts less than 500 fs is an indication of the AC Stark
effect. This short-lived AC Stark feature in Fig.~\ref{GaNdt}b is observed
in all the samples as the very fast increase and decay of the
spectrally-integrated DT for below band gap excitations (e.g.
Fig.~\ref{GaNIntDT}). The residual change in the DT which persists as long
as 400 ps is due to real excitation of carriers by the two-photon
absorption of the intense pump pulse (300 $\mu$J/cm$^2$).

At longer delays, induced absorption is observed at the GaN energy for
both above and below band gap excitations. The observed slow relaxation is
due to decaying remnant carriers and excitons at the GaN band edge. The
rate of this decay is observed to be slower than 300 ps, which is
consistent with the recombination lifetimes obtained for
GaN.\cite{MuthAPL1997,JursenasAPL2001}

\subsection{Differential transmission of an In$_{0.05}$Ga$_{0.95}$N epilayer}

The PL emission from the In$_{0.05}$Ga$_{0.95}$N epilayer
(Fig.~\ref{MQWPLPLE}) occurred at $\sim$3.20 eV for cw excitation
(100W/cm$^2$) from the HeCd laser (3.82 eV). This is slightly lower than
the band gap values for In$_{0.05}$Ga$_{0.95}$N in the literature,
constrained between $3.220$ eV to $3.224$
eV.\cite{wetzelAPL98,mccluskeyAPL98} TI-PL is only observed for above
bandgap excitation, never for below bandgap excitation. The cw-absorption
(Fig.~\ref{MQWPLPLE}) suggests an In$_{0.05}$Ga$_{0.95}$N band edge of
3.26 eV. When the broadening of the band edge is taken into account, an
effective band edge of 3.22 eV is obtained from a sigmoidal
fit\cite{MartinAPL1999}, giving a 20 meV Stokes' shift. Since the
piezoelectric (PZE) field in the InGaN epilayer is already reduced
remarkably with Si doping, the band edge changes very little with pulsed
laser excitation. Thus, screening of the PZE field does not produce a
remarkable blueshift. This is verified by pulsed absorption measurements
which suggest an In$_{0.05}$Ga$_{0.95}$N band edge of $\sim$3.26 eV,
similar to the cw-absorption.

\begin{figure}
\centerline{\resizebox{8cm}{!}{\hbox{\includegraphics{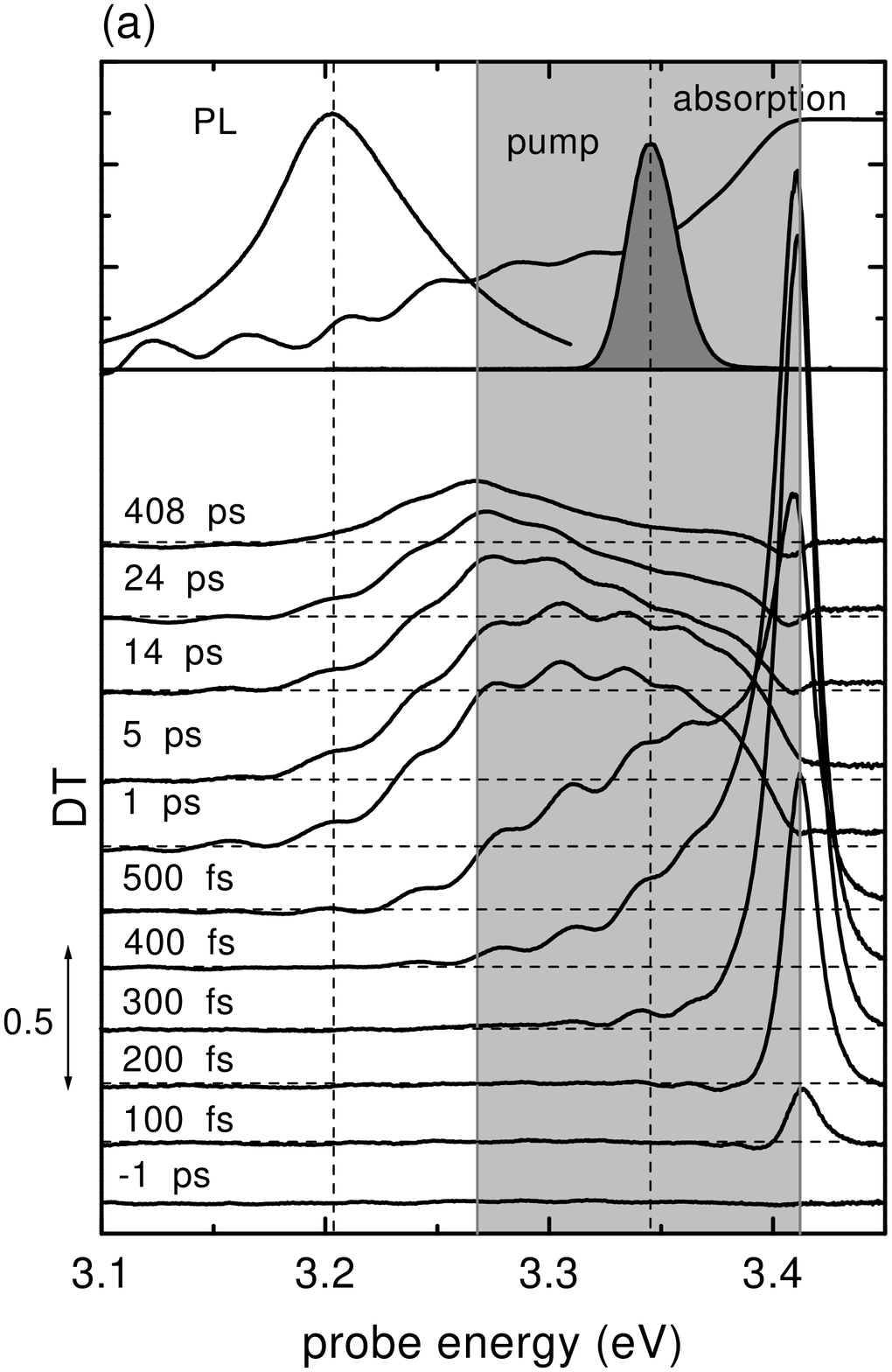}
\includegraphics{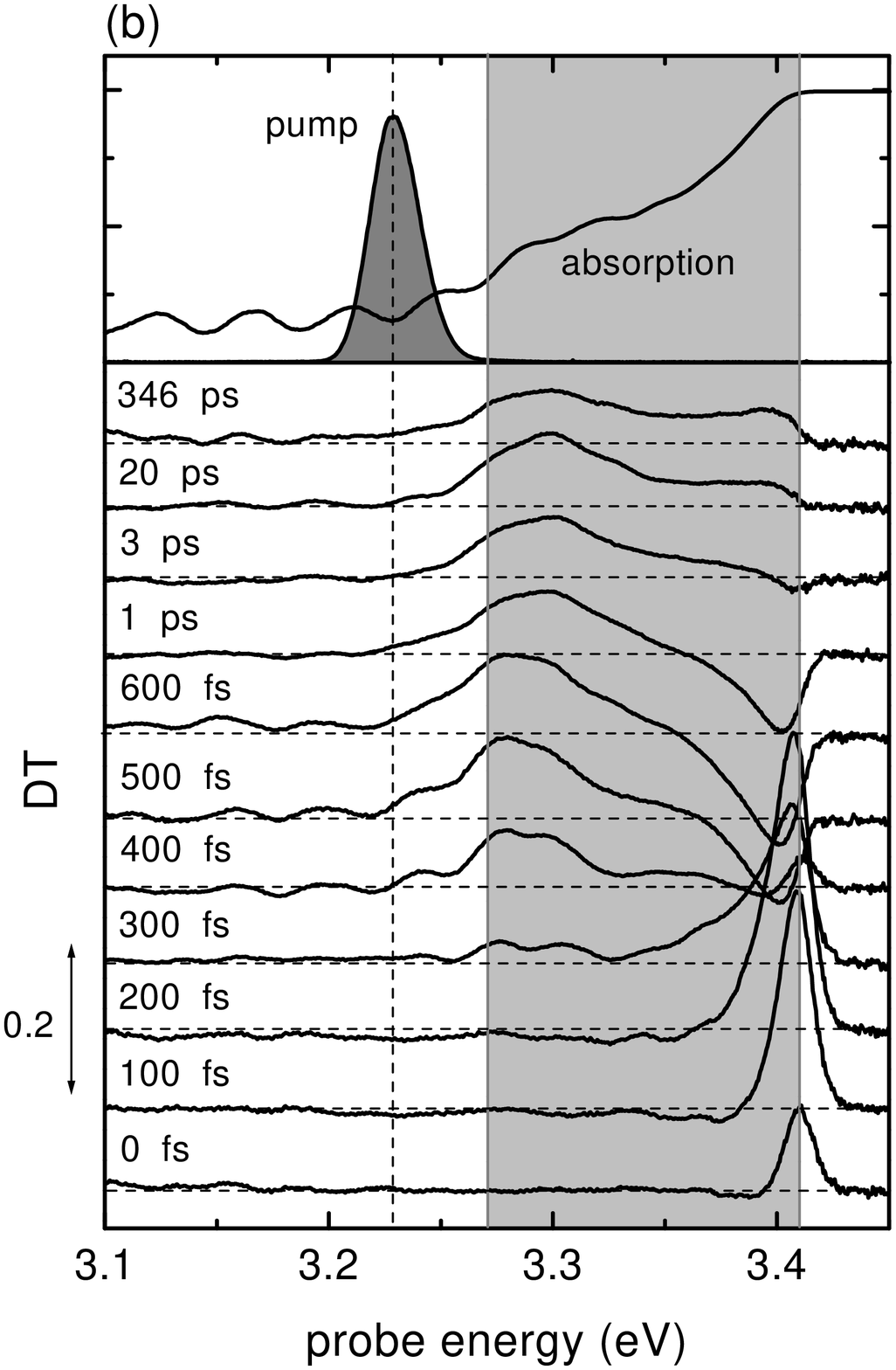}}} }
\caption{Spectrally-resolved TRDT for above (a) and below (b) bandgap
excitations of the InGaN epilayer sample at various delays for an
excitation density of 300 $\mu$J/cm$^2$. The top graphs show the pump
spectrum, the TI-PL, and the pulsed absorption. The shaded regions
indicate the states between GaN and InGaN band edges.} \label{InGaNdt}
\end{figure}

\begin{figure}
\centerline{\resizebox{8cm}{!}{\hbox{\includegraphics{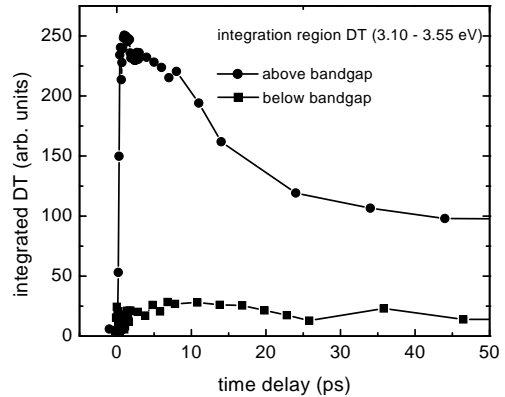}}}}
\caption{Time-evolution of the spectrally-integrated TRDT for above
(circles) and below(squares) bandgap excitations of the InGaN epilayer
sample at an excitation density of 300 $\mu$J/cm$^2$. Integration region
was 3.10 eV - 3.55 eV.} \label{InGaNIntDT}
\end{figure}

The spectrally-resolved TRDT data for the InGaN sample is shown in
Fig.~\ref{InGaNdt}. Pump wavelengths for Fig.~\ref{InGaNdt}a and
Fig.~\ref{InGaNdt}b were above (3.35 eV) and just below (3.23 eV) the
InGaN bandgap, respectively. Before considering carrier dynamics in the
InGaN layer, it is useful to reconsider carrier dynamics in the GaN layers
in this sample. The pump energies are 60 meV, and 180 meV below the GaN
bandgap, respectively. As in the below bandgap data for bulk GaN
(Fig.~\ref{GaNdt}b), it is first observed that the transmission increases
from carrier generation and AC Stark effect in the GaN, reaching a maximum
in $\sim$ 300 fs at the GaN band edge (3.41 eV) for both pump energies. AC
Stark effect is observed in the GaN layers of the InGaN epilayer sample in
a manner similar to the below band gap excitation of the GaN sample. The
magnitude of the AC Stark effect is smaller for the larger detuning of 190
meV in Fig.~\ref{InGaNdt}b. Real carriers are also generated at the GaN
band edge due to two-photon absorption, but unlike the GaN sample they can
quickly ($<$1 ps) decay into lower InGaN states through carrier-LO phonon
and carrier-carrier scattering processes. This makes the distinction
between the two-photon absorption and the AC-Stark effect more difficult.
The small decreases (above bandgap) or increases (below bandgap) in
absorption at the GaN band edge (3.41 eV) after 5 ps are due to the
remnant carriers and excitons at the GaN interface and trap states. The
two-photon induced carrier absorption feature in bulk GaN
(Fig.~\ref{GaNdt}b) remains visible after 408 ps with no remarkable
reduction in its amplitude. This contrasts with the more complex induced
absorption and transmission in the GaN cap and buffer layers of the InGaN
sample (Fig.~\ref{InGaNdt}b) which varies with excitation energy and time
but remains for at least 400 ps. The differences can be attributed to
carrier relaxation from the GaN layers to the InGaN epilayer and the
reduced role of SPE in the GaN layers and trap states.

Regarding carrier dynamics in the InGaN layer, SE appears at 3.21 eV, near
the InGaN band edge and slightly on the blue side (10 meV) of the PL, for
excitation above (3.35 eV) the InGaN bandgap. Since the SE peak is broad
and very close to the main PL peak, the onset of SE is observed as the
emergence of a narrower linewidth PL whose strength increases with
increasing pump intensity. From PL measurements of spectrally-integrated
intensity and linewidth, a pump threshold of $\sim$80 $\mu$J/cm$^2$ is
obtained. Spectrally-integrated TRDT data for above bandgap, above
threshold (300 $\mu$J/cm$^2$) excitation shows a fast decay during the
first 14 ps due to carrier removal through SE, followed by a much slower
relaxation through SPE after the carrier density decreases below the SE
threshold (Fig.~\ref{InGaNIntDT}). The fast rising edge of the
spectrally-integrated dt for both excitations include the contribution
from the AC Stark effect at the GaN band edge, as observed in
Fig.\ref{GaNIntDT}.

The fast relaxation feature of the spectrally-integrated TRDT is
spectrally-resolved in Fig.~\ref{InGaNdt}a as a redistribution of the
carriers. The broad distribution of carriers becomes clearly visible after
400 fs as a bleaching of the InGaN photo-absorption. The carrier
distribution extends from the GaN band edge to the InGaN band edge and
reaches a maximum in 1 ps at $\sim$ 3.30 eV. Afterwards, the blue edge of
this bleaching, $\sim$ 70 meV above the InGaN band edge, is observed to
decay rapidly ($<$ 14 ps) while the red edge (3.18-3.27 eV) remains almost
constant. The red shift of the bleaching arises from carrier cooling, and
the peak reaches the In$_{0.05}$Ga$_{0.95}$N band edge (3.26 eV) by 14 ps
when SE ceases. Thus the fast decay feature of the spectrally-integrated
TRDT is due to the fast removal of the carriers at the InGaN band edge
through SE, while carriers at higher energies relax down to refill the
lost carriers. After the number of carriers is reduced below the SE
threshold, only SPE remains.

By contrast, for the below bandgap excitation (3.23 eV), SE is not
observed, and the DT signal, shown in Fig.~\ref{InGaNdt}b, does not
exhibit SE-related features. The number of induced carriers is smaller,
and the DT signal shows them narrowly distributed close (3.27 eV) to the
band edge, reaching a maximum in 600 fs. Of particular noteworthiness is
the absence of the SE-mediated fast decaying blue edge seen in above
bandgap excitation. During the first 3 ps, there is also a more complex
absorption change at the GaN band edge due to a combined effect of the AC
Stark effect and the two-photon absorption. Induced absorption is observed
after 400 fs until 3 ps, and increased transmission at the GaN energy
persists for as long as 346 ps due to carriers and excitons in GaN trap
states.

Spectrally-integrated DT data for both excitations show an additional
decay component in the InGaN with a much larger time constant of several
100 ps. Spectrally-resolved TRDT reveals that carriers decay very slowly,
and the carrier distribution continues to narrow, after 24 ps in above
bandgap excitation (Fig.~\ref{InGaNdt}a) and after 600 fs in below bandgap
excitation (Fig.~\ref{InGaNdt}b). The slow decay of these cooled carriers
is due to radiative recombination-induced SPE. From the data, the
radiative recombination time in this InGaN layer is estimated to be
0.98$\pm$0.08 ns, a value consistent with the reports in the literature.
\cite{ChichibuJVST1998,KoronaMSEB2002} Other TRDT measurements on InGaN
epilayers \cite{ChoiPRB2001-2,OmaePSSA2002} show similar slow decay
behavior as in In$_{0.05}$Ga$_{0.95}$N epilayer investigated here.

\subsection{Differential transmission of an InGaN MQW}

\subsubsection{Investigation of Sample Structure}

As seen in Fig.~\ref{MQWPLPLE}, the cw-PL for the MQW sample shows a
single emission centered at $\sim$3.01 eV associated with the confined QW
minimum. The corresponding PLE and cw-absorption measurements indicate the
In$_{0.05}$Ga$_{0.95}$N barrier energy is $\sim$3.23 eV. This barrier
energy is 30 meV below the band edge energy obtained for the
In$_{0.05}$Ga$_{0.95}$N epilayer because the barrier composition is
slightly different, most probably due to In incorporation into the
barriers from the wells.

Simple 1-D calculations predict the minimum confined QW energy level
within 60 meV of the measured 3.01 eV, given reasonable values for
conduction:valence band offsets (20:80 TO 80:20) and bowing parameter
(1.75 eV)\cite{wetzelAPL98}. The difference between the measured and
calculated values is satisfactory given the large but uncertain PZE field
strength arising from lattice mismatch-induced strain between the layers.
\cite{TakeuchiAPL1998}

PLE at room temperature (Fig.~\ref{MQWPLPLE}) shows clear edges at the GaN
and the In$_{0.05}$Ga$_{0.95}$N barrier energies consistent with the
cw-absorption. The contribution from the confined QW states is not
observed in cw-absorption. However, the logarithmic plot of the PLE
signals at 303 K and 83 K (Fig.~\ref{MQWPLPLE} inset) shows the broad QW
edge at $\sim$3.11 eV and $\sim$3.14 eV, respectively. At room temperature
the emission is 100 meV lower than this PLE edge, but a sigmoidal fit to
this QW edge gives an effective band edge of 3.06 eV, and a Stokes' shift
of 50 meV. Shifts even larger than 300 meV have been observed for higher
In composition MQWs under cw
excitation.\cite{ChichibuMSEB1999,MartinAPL1999}

High excitation density TI-PL from the MQW sample produced a blueshift as
large as 90 meV from the low intensity cw-PL peak. Similar to the SE
feature of the InGaN epilayer, the MQW SE emission peak is not observed as
a narrow feature on the main PL. Instead, just above the threshold
($\sim$95 $\mu$J/cm$^2$), the emission peak narrows and blueshifts 20 meV.
As excitation intensity increases, this narrowed emission continues to
blueshift. This blueshift is primarily due to increased screening of the
PZE field and to lateral carrier confinement in the MQW. As in the InGaN
epilayer, the strain-induced PZE field of the MQW is decreased through Si
doping and through screening by the large number of carriers induced by
the pulsed laser.\cite{ChichibuMSEB1999} MQW lateral carrier confinement
arises from the formation of In-rich quantum dot-like regions in the QWs
which grow with increasing In composition.\cite{ChichibuJVST1998} Such
inhomogeneities have been observed for InGaN MQWs having even less than 15
$\%$ In.\cite{ChichibuMSEB1999,ChichibuAPL1998-2} Carrier localization due
to inhomogeneities is also expected to decrease with Si doping in the
barriers.\cite{ChoAPL1998} As the size of In-rich regions grow, the
confinement within large dots is reduced and the Stokes' shift increases.
The degree of confinement added to the screening of the PZE field make the
blueshift in the MQW TI-PL larger than it is for the InGaN epilayer
sample.

\subsubsection{Spectrally-integrated and -resolved TRDT}

\begin{figure}
\centerline{\resizebox{8cm}{!}{\hbox{\includegraphics{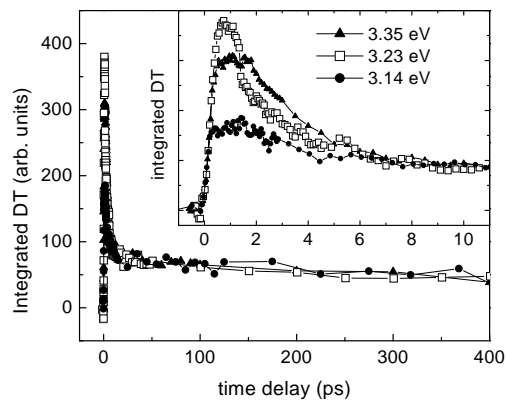}}}}
\caption{Time-evolution of the spectrally-integrated TRDT for above-
(triangles), at- (squares) and below-barrier (circles) bandgap excitations
of the InGaN MQW sample at an excitation density of 300 $\mu$J/cm$^2$.
Integration region was 2.95 eV - 3.40 eV. Inset shows the fast decay
components of the data.} \label{MQWINTdt}
\end{figure}

To examine the relaxation of the total number of carriers, DT signals are
integrated over the spectrum 2.92 - 3.41 eV for excitation energies above
(3.35 eV), at (3.23 eV), and below (3.14 eV) the barrier
energy.\cite{integratedDT} As seen in Fig.~\ref{MQWINTdt}, all three
spectrally-integrated TRDT signals show a fast decay at early times,
followed by a very slow relaxation. The fast components are observed to
decay in less than 10 ps. Spectrally-integrated DT data were fit by a
bi-exponential decay function, $Ae^{-t/\tau_1}+Be^{-t/\tau2}$, where
$\tau_1$ and $\tau_2$ are the decay times for the fast and the slow
decaying components, and A and B are the corresponding amplitudes. A
fractional strength value ($f=A/(A+B)$) is defined to observe the relative
strength of the fast decaying component.

\begin{figure}
\centerline{\resizebox{8cm}{!}{\hbox{\includegraphics{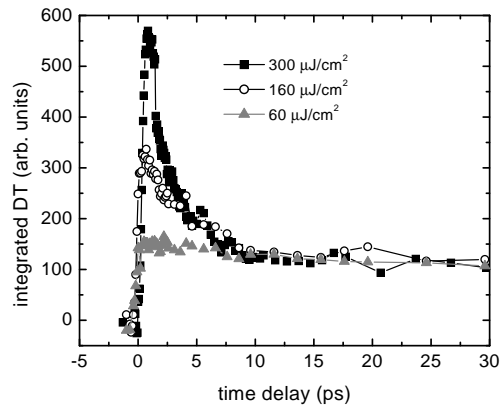}}}}
\caption{Time-evolution of the spectrally-integrated TRDT for at-barrier
excitation of the InGaN MQW sample for different pump power densities.
Data were normalized to the slow decaying amplitude. With decreasing power
density the SE-mediated fast decaying component is observed to disappear.}
\label{IntDTPower}
\end{figure}

The fractional strength ($f$) of the fast decaying component is largest
for at-barrier excitation, and is larger for above-barrier excitation than
for below-barrier excitation (Inset, Fig.~\ref{MQWINTdt}). As with the GaN
and InGaN epilayers, the fast decay in the MQW sample is caused by the
accelerated relaxation of carriers through SE. However, SE is observed
even for below-barrier excitation of the MQW sample. For at-barrier energy
excitation (SE threshold $\sim$ 95 $\mu$J/cm$^2$), the magnitude of the
fast feature is observed to decrease ($f$=0.74, 0.65 , 0.26) and its decay
constant slowed (2.6, 4.6, and 13.5 ps) as the excitation density
decreased (300, 160, and 60
$\mu$J/cm$^2$)(Fig.~\ref{IntDTPower}).\cite{SEbelowThreshold} For a given
excitation density (300 $\mu$J/cm$^2$), the fast feature decay times for
at-, above-, and below- barrier excitations increased (2.6, 2.6, and 3.7
ps, respectively)(Fig.~\ref{MQWINTdt}), and the magnitude of the feature
decreased ($f$=0.74, 0.73, 0.56), in a manner consistent with the strength
of the PLE measured at the respective excitation energies
(Fig.~\ref{MQWPLPLE}).

SE decay times faster than 10 ps have been observed for similar MQW
structures at room temperature and at 2
K.\cite{SatakePRB1999,KawakamiJPCM2001} In one study, a three-level rate
equation model was also developed, consisting of 2D QW, 3D barrier, and
ground (recombined) states.\cite{SatakePRB1999} This model suggested that
the fast relaxation was due to carriers decaying from the 3D states to
refill the 2D states emptied by SE. The fact that higher energy 3D states
supply carriers to lower energy states undergoing SE has been confirmed in
the GaN (Fig.~\ref{GaNdt}a) and InGaN (Fig.~\ref{InGaNdt}a) data presented
above. As will be discussed below, a similar process occurs in the MQWs,
in which 3D states supply carriers for the saturated 2D states undergoing
SE. Thus, SE reduces the total number of carriers in a similar manner for
all samples studied.

\begin{figure}
\centerline{\resizebox{8cm}{!}{\hbox{\includegraphics{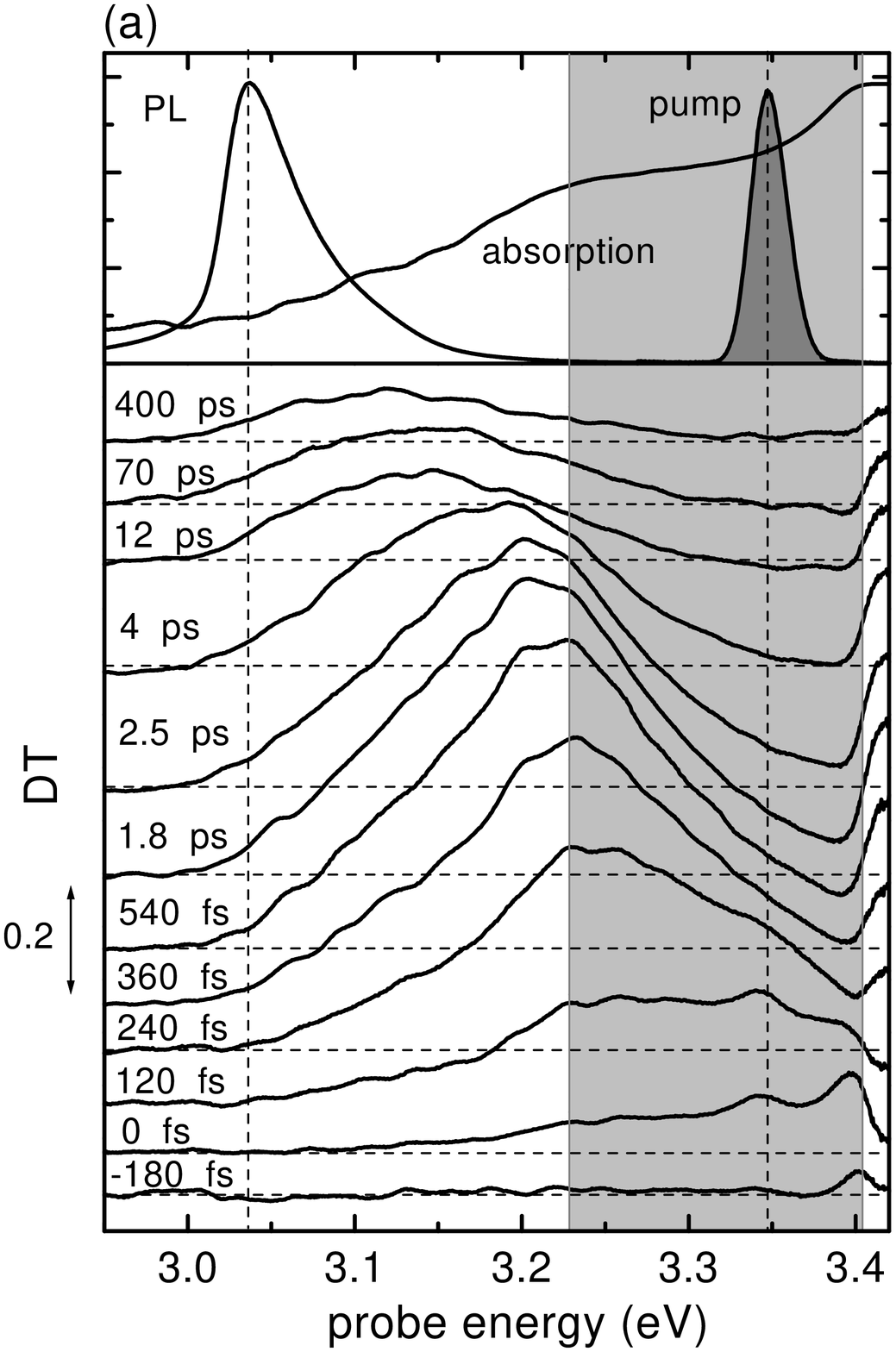}
\includegraphics{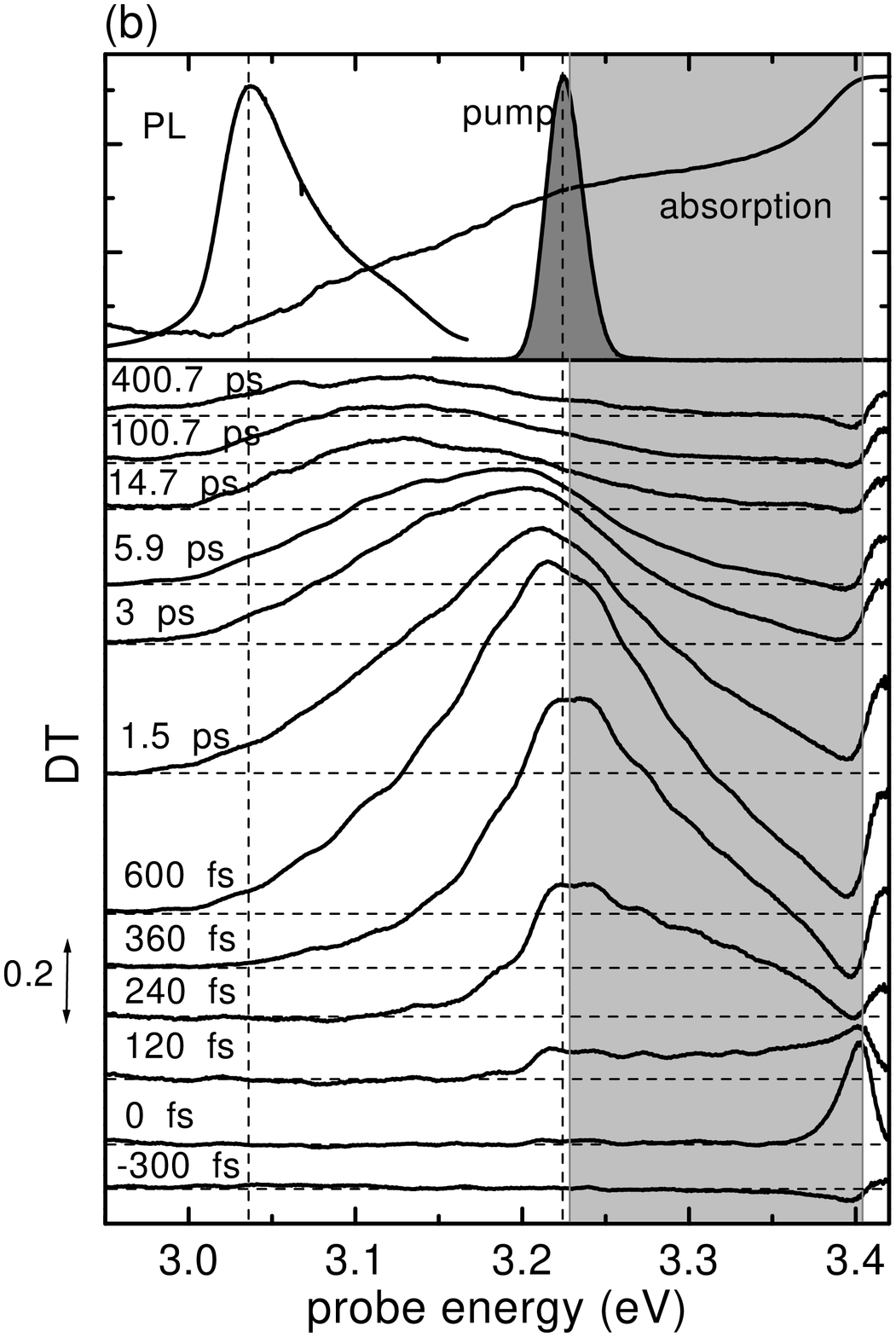}}} }
\centerline{\resizebox{8cm}{!}{\hbox{\includegraphics{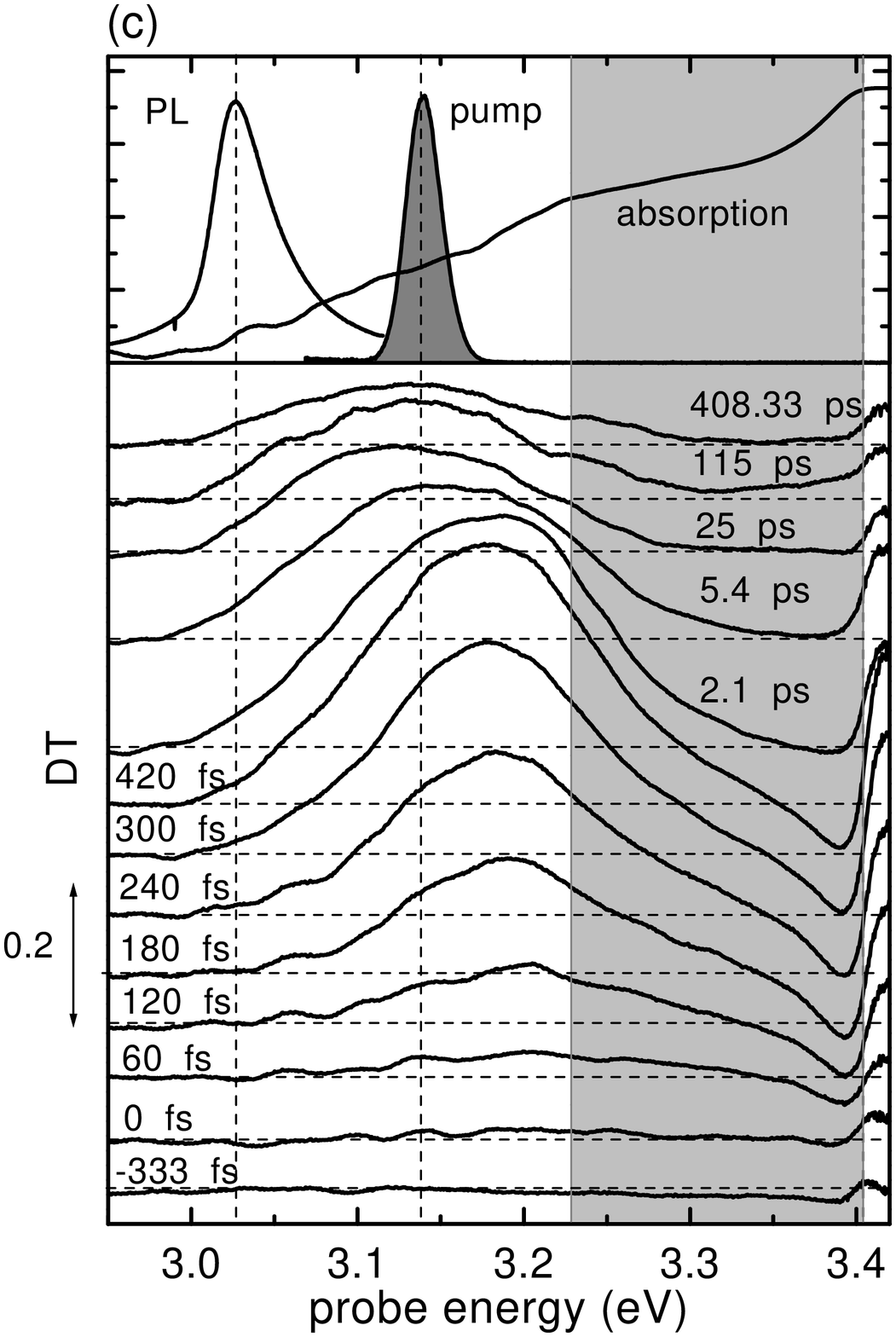}
\includegraphics{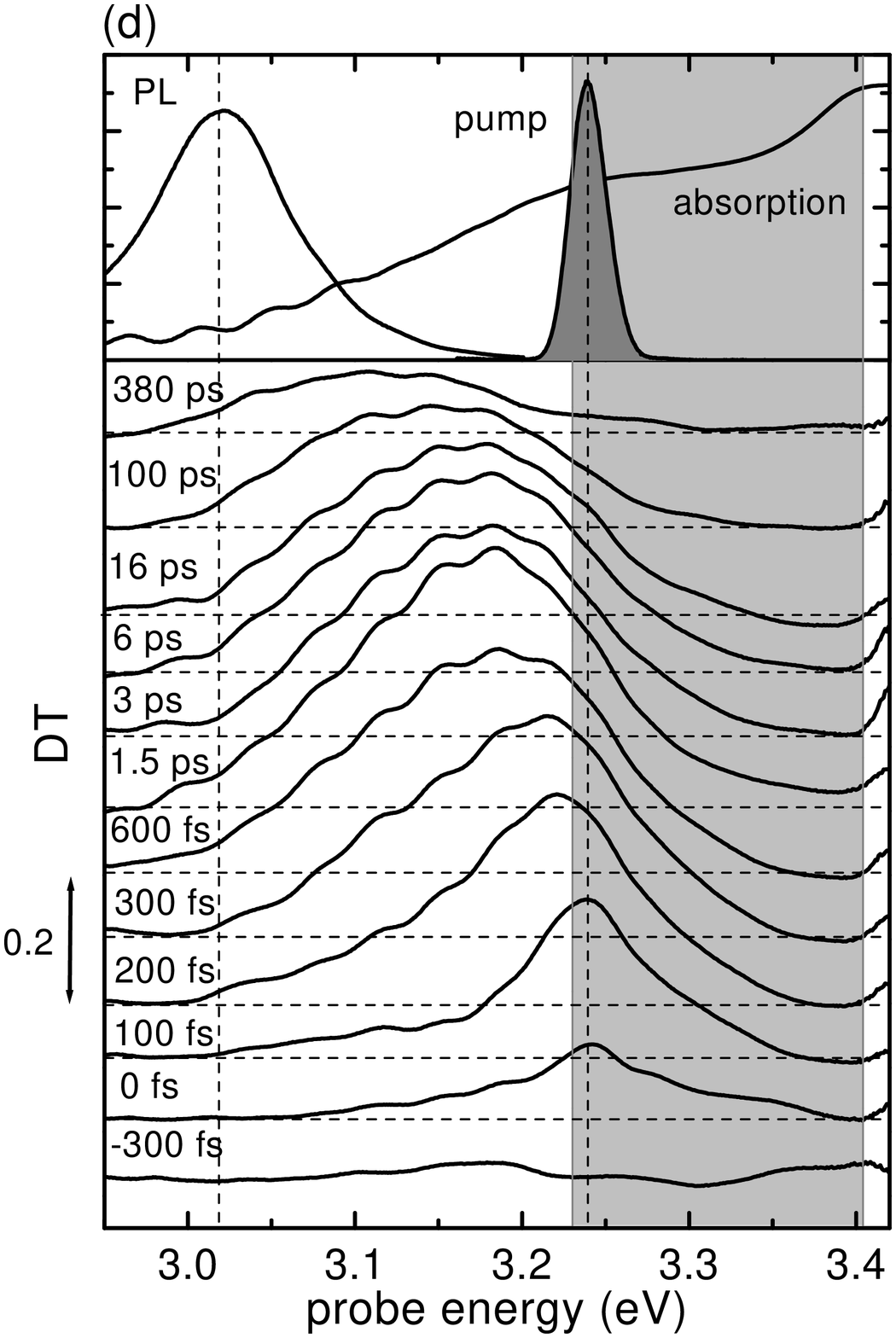}}} }
\caption{Spectrally-resolved TRDT for above- (a), at- (b), and
below-barrier (c) energy excitations at various delays for an excitation
power density of 300 $\mu$J/cm$^2$. Pulsed absorption shows a clear
absorption edge at $\sim$ 3.23 eV, and the shaded regions indicate the
states between the barrier and the GaN band edges. (d) shows the TRDT data
for at-barrier excitation with a lower power density of 60 $\mu$J/cm$^2$.}
\label{MQWdt}
\end{figure}

To elucidate the carrier redistribution and relaxation processes,
spectrally-resolved TRDT data for the InGaN MQW sample for various delays
at different pump energies above, at, and below the barrier energy are
shown in Fig.~\ref{MQWdt}. The fast decaying AC-Stark and the two-photon
absorption feature at the GaN energy (3.41 eV) is observed at all pump
excitation energies and behaves analogously to the below GaN bandgap
excitation for the GaN and InGaN epilayer samples discussed above. The
strength of this feature decreased both with decreasing pump intensity and
increasing detuning.

In further analogy with the InGaN epilayer, a broad bleaching is observed
in the InGaN MQW barrier region (3.13 - 3.30 eV) for all excitation
energies. The peak of the carrier distribution is observed at the 3.23 eV
barrier energy. For above- and near-bandgap excitation in
Fig.~\ref{MQWdt}a and Fig.~\ref{MQWdt}b, the blue edge of this carrier
distribution rises faster (300 fs, 360 fs) than the red edge (540 fs, 560
fs) in the QWs.  This is reminiscent of the carrier redistribution for
above bandgap (3.35 eV) excitation in the InGaN epilayer
(Fig.~\ref{InGaNdt}a).

Likewise, the blue edge of the MQW carrier distribution, centered at the
3-D barrier band edge (3.23 eV), is observed to decay much faster than the
red edge (blue:10ps, red:$>$100ps). As in the InGaN epilayer
(Fig.~\ref{InGaNdt}a), this initial fast decay of the blue edge
corresponds to the SE-related fast decay in the spectrally-integrated
TRDT. However, the carriers cool down to the barrier band edge more
quickly ($<$1 ps) than they did for above bandgap excitation of the InGaN
epilayer, resulting in a very broad (170 meV) carrier distribution
extending from the GaN edge down into the QWs. This difference between
epilayer and MQW barrier carrier relaxation arises because capture into
the QWs removes carriers from the InGaN barrier much faster than
recombination removes them from the InGaN epilayer.\cite{OzgurAPL2000}

The role of SE-mediated carrier relaxation is most clearly seen by
comparing the at-bandgap excitation TRDT data above (300 $\mu$J/cm$^2$)
(Fig.~\ref{MQWdt}b) and below (60 $\mu$J/cm$^2$) (Fig.~\ref{MQWdt}d)
threshold.  In the first 1 ps, the larger, asymmetric carrier distribution
for the above threshold data is obvious, indicating the greater carrier
concentration in the barriers and the role of SE in effectively depleting
the higher energy barrier states faster than the band edge barrier states.
Although carrier capture is apparent, a large percentage of carriers
remain in the barrier region while SE is operating. By contrast, the below
threshold TRDT data indicates a more symmetric carrier distribution whose
relaxation rates are much less sensitive to carrier energy and whose peak
drops below the barrier band energy in only 300 fs.  By 1.5 ps, the below
threshold carrier distribution is fully 50 meV below that of the above
threshold case, and the barriers are virtually depopulated.  It may be
surmised that SE assists in the carrier capture process by quickly
depopulating QW states which are refilled by captured barrier band edge
carriers which in turn are refilled by higher energy carriers in the
barriers.  The normal cooling of the carrier distribution observed below
threshold is altered above threshold by SE through the continued emptying
of energetically accessible states near the barrier band edge.  By 6 ps
when SE is over, however, the distributions appear quite similar and
exhibit the same relaxation behavior thereafter.

\begin{figure}
\centerline{\resizebox{8cm}{!}{\hbox{\includegraphics{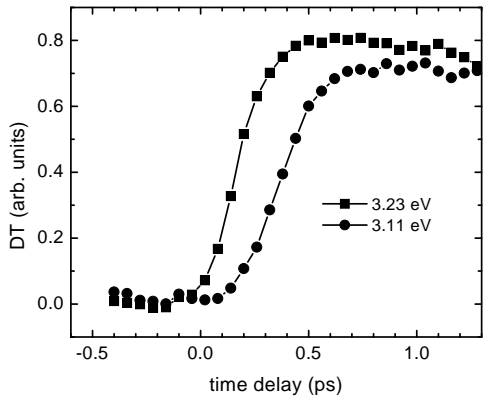}}}}
\caption{Normalized DT at barrier (3.23 eV, squares) and the QW band
edge(3.11 eV, circles) energies for the at-barrier energy excitation (300
$\mu$J/cm$^2$). The rise of the DT at QW energy is slower than the rise of
the signal at the barrier energy, indicating electron capture into the
confined QW states.} \label{Capture}
\end{figure}

In addition to elucidating the action of SE and carrier redistribution,
Fig.~\ref{MQWdt} reveals the mechanism of QW carrier capture. Due to their
higher density of states, holes are expected to be captured from the 3-D
barriers to the 2-D QW levels much faster than the
electrons.\cite{MansourJAP1995} Pulse widths of 100 fs used here limit our
observations to the electron capture. A significant number of carriers are
observed at and above the barriers after 240 fs for above- and at-barrier
excitations, while carriers are just beginning to appear in the QWs
(Fig.~\ref{MQWdt}b). The carrier distribution at the barrier band edge
(3.23 eV) peaks in $\sim$ 0.5 ps, while the carrier distribution at the
QWs (3.11 eV) reaches its maximum in $\sim$ 0.8 ps (Fig.~\ref{Capture}).
This delay of 0.3 ps confirms that the carriers are reaching the barriers
faster than they arrive at the QWs, an indication of the electron capture
process. This result is consistent with previous measurements of electron
capture time (340-510 fs) using degenerate TRDT
spectroscopy.\cite{OzgurAPL2000}

In addition, degenerate TRDT showed optimal emission efficiency and most
efficient carrier capture occurred when the carrier injection was within
$\pm$ 40 meV of the 3.23 eV barrier energy. \cite{OzgurAPL2000} Similar to
the cw-absorption measurements (Fig.~\ref{MQWPLPLE}), the pulsed
absorption spectrum in Fig.~\ref{MQWdt} clearly shows an 80 meV-wide peak
centered at 3.23 eV. This broad absorption suggests potential fluctuations
in the barriers due to the compositional inhomogeneities. The
spectrally-integrated DT data obtained here verifies that the most
efficient carrier capture occurs for pump energies near the barrier
energy.

Measurements by Satake \emph{et al.}\cite{SatakePRB1999} also show a broad
distribution of carriers around the barrier energy and SE-related decays
faster than 5 ps in the spectrally-resolved DT. However, the QW capture
rate could not be observed.  Other studies on InGaN-based laser diode
structures by Kawakami \emph{et
al.}\cite{KawakamiAPL2000,KawakamiJPCM2001} also observed the same fast
carrier accumulation at the barriers and much slower SPE decay from the
QWs. It is not clear if those samples, excited at power densities above
the SE threshold, show the same initial fast decay feature in the
spectrally-integrated DT signals as seen here. However, carrier capture to
the lower energy localized QW states appears to occur in less than 1 ps,
similar to our measurements. Surprisingly, a sizable carrier population
remains at barrier energies for at least 200 ps in their samples, which
they claim is due to the capture of carriers in a high lying, nearly
delocalized QW state. However, this feature bears some similarity to the
long-lived excitonic relaxation in GaN interfacial traps observed in our
InGaN TRDT data (FIG.~\ref{InGaNdt}). By contrast, the broad, quickly
decaying distribution of carriers observed in our MQW sample is probably
due to rapid carrier capture from the broad absorption at the barrier
energy (Fig.~\ref{MQWdt}).

For an excitation energy (3.14 eV) below the barriers, total carrier
generation is smaller, and the peak carrier distribution appears at a
slightly lower energy than the other two excitation energies
(Fig.~\ref{MQWdt}c). Unlike below-band edge (3.23 eV) excitation in the
InGaN epilayer, the MQW carrier distribution is observed to peak 50 meV
below the barrier edge. Nevertheless, both the InGaN and MQW carrier
distributions reach their maximum in a very short time ($\sim$400 fs), and
the rise times for both blue and red edges of the respective carrier
distributions (Fig.~\ref{InGaNdt}b and Fig.~\ref{MQWdt}c) are similar.
This behavior contrasts with the above- and at- bandgap excitation for the
MQW in which the blue side rises faster than the red side. As observed in
Fig.~\ref{MQWINTdt}, the fast decaying component of the
spectrally-integrated DT for the below-barrier excitation is smaller than
for the other two excitations, indicating a weaker carrier decay through
SE. Accordingly, $f$ is very small for the spectrally-integrated DT, and
the blue edge decay of the TRDT signal in Fig.~\ref{MQWdt}c is not
noticeable.

The fast decaying component in the spectrally-integrated TRDT signal
disappears as the carrier density falls below the SE threshold and
emission goes from SE to
SPE.\cite{KawakamiAPL2000,SatakePRB1999,KawakamiJPCM2001} After SE ends (5
- 10 ps), the carriers at the QW band edge recombine only through SPE, and
carriers at higher energies relax at a similar rate to replace those lost
at the QW band edge. For all excitations this is observed as the redshift
of the bleaching. The carrier distribution cools and redshifts to the QW
band edge ($\sim$ 3.11 eV) after 400 ps. In addition to carrier
recombination in the QWs, the redshift of the carrier distribution is also
partly due to the reduced screening of the PZE field as carrier density
decreases through SPE.\cite{KawakamiAPL2000} An exponential fit to the
spectrally-resolved data at 3.11 eV gives a decay constant of
0.66$\pm$0.06 ns, consistent with the 0.69 ns value from previous
degenerate TRDT measurements.\cite{OzgurAPL2000} The slow decay constant
from the spectrally-integrated TRDT is slightly smaller (0.54$\pm$0.07
ns), but agrees satisfactorily when the carrier redistribution through
different channels is taken into account.

\section{Conclusions}

In summary, non-degenerate TRDT spectroscopy is performed on a bulk GaN
layer, an InGaN epilayer, and an InGaN MQW. All the samples were observed
to have SE features for excitation densities above a pump threshold
density of $\sim$100$\mu$J/cm$^2$. Spectrally-integrated TRDT data showed
the effects of SE-mediated decay, which occurred in $<$10 ps for above
band gap excitations. These fast decays are accompanied by carrier
relaxation from higher to lower energy states which are emptied in turn by
SE. After the total carrier density is reduced below threshold, SE
disappeared and a slow relaxation through SPE is observed. For GaN, a
spectrally narrow absorption of carriers decayed by SE at the band edge in
6 ps. A broader carrier distribution is observed for the InGaN sample
which narrowed and decayed to the InGaN band edge through SE in $<$14 ps.
Carrier recombination through spontaneous emission was observed with a
time constant of approximately 1 ns.

Photo-excitation above, at, and below the barrier energy for the MQW
sample showed more complex relaxation pathways in the QWs.
Spectrally-integrated TRDT signals for above- and at-barrier bandgap
energy excitations demonstrated SE as a fast decaying component in the
first 10 ps. The strength and fast decay times of the SE feature were seen
to vary as a function of excitation energy and density.  For a given
excitation density, the decay times varied inversely with the PLE
magnitudes for the respective excitation energy. SE was observed to
depopulate the higher energy barrier states faster than the lower energy
barrier states through a process of cascaded refilling. Once SE ended,
carrier capture and spontaneous emission from carrier recombination cools
the carrier distribution into the wells and toward the lowest energy QW
state. The wavelength non-degenerate TRDT data at QW and barrier energies
provided sub-ps values for the electron capture time, and $\sim$660 ps for
the recombination time. These values agree well with previous degenerate
TRDT measurements on the same MQW sample.\cite{OzgurAPL2000}

\begin{acknowledgments}

The authors would like to thank Steven DenBaars, Stacia Keller and Amber
Abare from University of California, Santa Barbara for supplying the
samples, and Arup Neogi for helpful discussions. This work was supported
by the Army Research Office.

\end{acknowledgments}


\begin{references}
\bibitem{GaNabsorption}
This cooling is not observed in TRDT due to high absorption above the GaN
band edge.
\bibitem{integratedDT}
It should be noted that part of the DT signal at the GaN band edge is
excluded from the spectral integration to avoid obfuscation by the AC
Stark effect.
\bibitem{SEbelowThreshold}
Note that a tiny SE feature remains even below the threshold due to pump
and sample inhomogeneities.
\end{references}
\end{document}